**Band bending at interfaces between topological insulator Bi$_2$Se$_3$ and transition metals**

Weiguang Ye,[1] A. B. Pakhomov,[2] Shuigang Xu,[1] Huanhuan Lu,[1] Zefei Wu,[1] Yu Han,[1] Tianyi Han,[1] Yingying Wu,[1] Gen Long,[1] Jiangxiazi Lin,[1] Gu Xu,[2] Yuan Cai,[2] Lu-Tao Weng,[2] Ning Wang[1]*

[1]Department of Physics and the William Mong Institute of Nano Science and Technology, [2]Materials Characterization and Preparation Facility, the Hong Kong University of Science and Technology, Hong Kong, China

* Corresponding author, email address: phwang@ust.hk

Interfaces between exfoliated topological insulator Bi$_2$Se$_3$ and several transition metals deposited by sputtering were studied by XPS, SIMS, UPS and contact I-V measurements. Chemically clean interfaces can be achieved when coating Bi$_2$Se$_3$ with a transition metal layer as thin as 1 nm, even without capping. Most interestingly, UPS spectra suggest depletion or inversion in the originally n-type topological insulator near the interface. Strong band bending in the topological insulator requires careful material engineering or electric biasing if one desires to make use of the spin locking in surface states in the bulk gap for potential spintronic applications.

Integration of topological insulators (TI)[1,2,3] with transition metals (TM), either by doping the TI with TM ions[4,5,6] or interfacing the TI surface with a TM film[7,8,9,10] is interesting from the point of view of both the new science involved and potential applications. Doping combined with band structure engineering can lead to intrinsic ferromagnetism in the TI mediated by topological states which can exhibit quantum anomalous Hall effect[11] at low temperatures, in excellent agreement with the theoretical prediction.[12]

Interfacing of a TI with a ferromagnetic film has been proposed for device structures based on spin valve or spin transfer torque effects, employing the spin locking to orbital momentum in topological surface states. Observation of the spin-orbit torque in TI-TM bilayers[7,8] opens a route to new spin transfer torque random access memory (STT-RAM) designs with the in-plane switching current.

In order to utilize topological states of the TI in spintronic designs, one needs to achieve dominant conduction via these states. Typically as-prepared pure $Bi_2Se_3$ is n-doped due to native defects such as Se vacancies, with the Fermi energy in the bulk conduction band.[13] The crossing point of the Dirac cone for topological states is in the gap of the bulk band structure, as shown in Fig. 4(c). One can achieve conduction primarily via surface states by appropriate doping or by electrical biasing, that is by tuning the position of the Fermi level to the bulk gap. However one important question which needs to be addressed is what happens to bulk states of the TI near the TM/TI interface, as well as to topological surface states at this interface. This question is not trivial as topological surface states are defined for an interface with vacuum or another dielectric. Recent publications show that topological surface states are tolerant to TM on the surface at least for relatively small coverage range (from 0.01 to 0.5 monolayer).[14,15,16,17] Another question is the occupation of electronic states of the TI which can be altered near metal contact. In a simplified consideration, the electron affinity of $Bi_2Se_3$ is about 4.45 eV[18] and the work functions of transition metals are close to 5 eV,[19] hence one can expect upwards band bending in the TI near the interface. Taking into account that the bandgap is just 0.3 eV, the effect of band bending can be relatively strong with $E_F$ moving from the conduction band to the bandgap or even to the valence band as $E_F'$ or $E_F''$ in Fig. 4 (d). Understanding this effect is practically important for the device design and material engineering. Shifts in the Fermi energy due to surface doping have been observed in Refs. [14,15] Downwards band bending was also observed in angle resolved photoelectron (ARPES) experiments on $Bi_2Se_3$ surfaces exposed to residual gas.[20] However, notice that for spintronic devices under consideration, [7,8,9,10] band bending at the interface with continuous and bulk, though thin (~nm), TM films are more important than the effect of individual dopant atoms. This problem has been less studied so far.

We investigated interfaces between $Bi_2Se_3$ and thin films of several TMs, including Co, Fe, Ni and Cr, directly by surface science techniques: x-ray photoelectron spectroscopy (XPS), secondary ion mass spectrometry (SIMS), and ultraviolet photoemission spectroscopy (UPS). The film deposition was done *ex-situ* by sputtering, which is the most typical industrial method for magnetic films. A combination of XPS and SIMS results shows that the interface is free from oxygen even when the metal film is as thin as 1 nm and without a protective capping layer, but some redox reactions may be observed. The most interesting result is the observation by UPS that while the Fermi level is located in the bulk conduction band on a freshly cleaved $Bi_2Se_3$ surface, it moves to the bandgap and valence band (sometimes rather deep under the bandgap top) when the TI is in contact with TM. This creates a depletion or inversion layer and results in a rectifying or highly resistive interface.

Bulk single crystalline $Bi_2Se_3$ was grown by the Bridgeman method.[21] These crystal platelets were cleaved and exfoliated using a sticky tape. These platelets were at least a few microns thick so the material could be considered bulk. After exfoliation these samples were placed in the sputtering chamber for metal deposition. We prepared three sets of samples with each metal: in the first set we deposited just 1 nm of TM on the fresh surface; in the second set the thickness of TM layer was 2 nm, capped with 3.5 nm of Pt. In the third set samples similar to those in the second set were additionally annealed at 300 C for 2 hours. For reference, we characterized a freshly cleaved surface and a sample oxidized in air for about a week. Here we show experimental results only for samples of set 1 and those reference samples. XPS and UPS measurements were performed on a Kratos Axis Ultra tool, and SIMS experiments on a IONTOF TOF-SIMS V spectrometer.

Fig. 1 shows the XPS spectra of Bi 4f for several samples with 1 nm TM coatings, along with the spectra for a freshly cleaved $Bi_2Se_3$ surface (Fresh) and a sample left in air for a week (Oxidized). The central peak corresponds to bismuth selenide. We notice Bi oxide both in the "oxidized" sample and in the sample coated with Cr; however Bi oxidation is weak with other TM coatings. Interestingly, some Bi metal (or reduced compared with the formal 3+ state) is seen in Cr- and Ni-coated samples revealing a redox reaction during the sputtering process. This agrees with the prior observation on $Bi_2Se_3$ surface with Fe adatoms.[15] Oxidation of the TM itself is observed too in these uncapped samples as they are exposed to air after sputtering.

It is seen in Fig. 2. (a-d) for Co, Fe, Ni and Cr coatings. We checked that annealing in vacuum does not increase the level of oxidation substantially, and a 3.5 nm Pt capping layer well protects the metals from oxidation.

The XPS probing depth (a few nm) is too large to resolve the layers at a 1-nm or less scale. In order to investigate the quality of these interfaces, we employed SIMS for depth profiling. These results are presented in Fig. 3 (a-d) for samples coated with 1 nm of Co, Cr, Fe and Ni. Though bismuth oxide is present in the SIMS results, it is interesting that it appears on the surface of TM rather than at the TI/TM interface. The interface is free from oxides even without Pt capping. The presence of Bismuth on the TM surface is most likely due to the secondary sputtering of Bi in the process of TM deposition. We conclude that the TM/TI interface is clean and mostly free from oxidation.

Most interesting results are obtained in UPS experiments where one can observe the effect of TM on the filling of valence states in the TI. Fig. 4 (a) shows the UPS spectra for $Bi_2Se_3$ (both freshly cleaved and oxidized in air for 1 week), and $Bi_2Se_3$ samples coated with 1 nm thick layers of several TMs. We are primarily interested in the spectrum at small binding energies (zoomed-in in Fig. 4 (b)). We associate the double peak measured on a fresh TI surface as due to electrons emitted from the valence and conduction bands of the TI (the freshly cleaved TI surface is assumed to be n-type). For illustration, schematic diagrams in Figs. 4 (c) and (d) show the expected band structure and density of states at a $Bi_2Se_3$ surface/interface, consistent with our results, where the bulk band gap is about 0,3 eV and the Dirac cone, describing surface states, has a crossing in the bulk band gap. [3] For a clean fresh surface, the Fermi energy is expected to lie in the conduction band, hence a double peak appears in the UPS spectrum in Fig. 4(b). (Notice that unlike ARPES [5,14,15,20], UPS does not have resolution required to resolve the surface state peaks). This double-peak feature disappears in the spectrum for oxidized surface. For TM-TI interfaces, the general trend is that the Fermi level is shifting down as shown in Fig. 4(d), resulting in a considerable reduction of the UPS intensity and a shift to higher electron energies in Fig. 4(b). Specifically, with Cr coating the Fermi energy apparently lies within the bandgap or close to the top of the valence band ($E_F$'), and for other TMs it is deeper in the valence band ($E_F$') so this UPS double peak is practically invisible for Co, Ni, Fe coated samples in Fig 4(b). Additionally, we performed a measurement on a Co coated sample after a deep *in-situ* sputtering with Ar to remove the TM coating completely; this sample is referred to as "sputtered $Bi_2Se_3$" in Fig. 4 a,b). The peak was restored on the sputtered surface as seen Fig. 4 (b), however the double-peak structure could not be resolved, probably due to surface disorder arising from Ar bombardment. This effect of shift and disappearance of the double-peak feature in the UPS spectrum of TM-coated samples can be interpreted as band bending in $Bi_2Se_3$ at the interface, which extends into $Bi_2Se_3$ much deeper than the probing depth of UPS, as expected at metal-semiconductor interfaces. It may lead either to depletion or inversion in n-doped $Bi_2Se_3$.

In both the depletion and inversion cases at TI-TM interfaces, one can expect that the interface should have non-Ohmic properties. We checked it by measuring I-V curves on TM/TI contacts. As patterning of a small metal contact on a flake was problematic, we chose to measure I-V curves on point contacts made on a freshly cleaved TI by a sharp tip coated with the TM of interest, specifically we used tungsten tips coated with Cr, Co, and Ni. Non-Ohmic behavior could be seen for all three coatings (Fig. 5). However, while the Cr-TI rectifying contacts were relatively easy to observe and behaved as Schottky barriers (Fig. 5(b)), in the cases of Co and Ni usually the initial contact was insulating, and required an electric breakdown before conduction could be measured. Figs. 5 (c) and (d) show examples of rare measurable contacts with Co and Ni, which also show breakdowns. The differences in the I-V behavior may be explained taking into account the UPS results: the down shift of the Fermi energy for Co and Ni is much stronger than that for Cr.

Results of our experiments show feasibility of fabrication of a chemically clean interface between $Bi_2Se_3$ and transition metals, even in the absence of capping layers. However, such interface alters electronic properties of the TI due to strong upwards band bending, or equivalently a down shift of the Fermi energy. Hence interface layers of the initially n-type TI can be depleted or inverted. This needs to be taken into account in the design of spintronic devices where the position of the Fermi level in the bandgap is crucial for the desired operation of the device. Actually, in some cases such band bending may have a positive effect, leading to the transformation of the doped n-type metallic material into a true topological insulator with conduction via topological surface states.

Acknowledgments

Financial support from the Research Grants Council of Hong Kong (Project Nos. 16302215, HKU9/CRF/13G) and technical support of the Raith-HKUST Nanotechnology Laboratory for the electron-beam lithography facility at MCPF (Project No. SEG_HKUST08) are hereby acknowledged. We thank Mr. Nick Ho of MCPF-HKUST for his assistance in the XPS/UPS measurements.

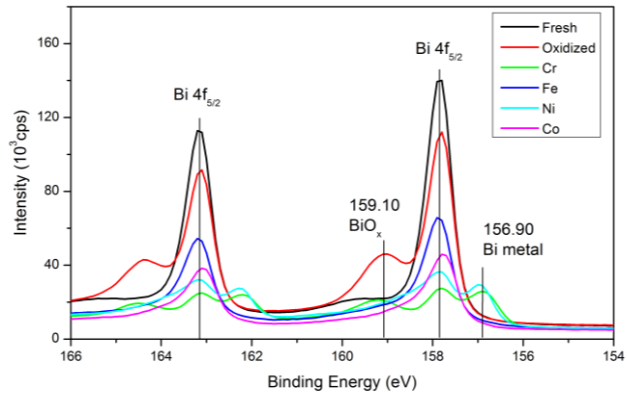

FIG. 1. XPS spectra of Bi 4f for different $Bi_2Se_3$ samples including a freshly cleaved $Bi_2Se_3$, an oxidized $Bi_2Se_3$ (1 week in air), and $Bi_2Se_3$ coated with Cr, Fe, Ni and Co. The peak at 159.10 eV indicates bismuth oxide. The peak at 156.90 eV is for Bi metal.

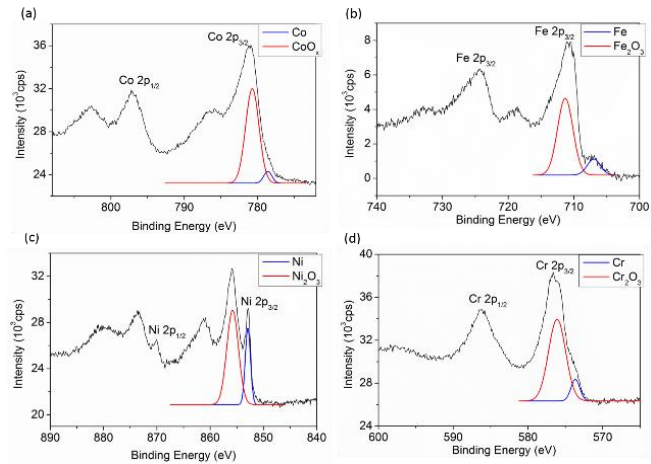

FIG. 2. 2p peaks of TMs in XPS spectra of $Bi_2Se_3$ coated with (a) Co, (b) Fe, (c) Ni and (d) Cr; the blue and red curves are fittings assuming Gaussian curve shapes for the TM and the metal oxide, respectively.

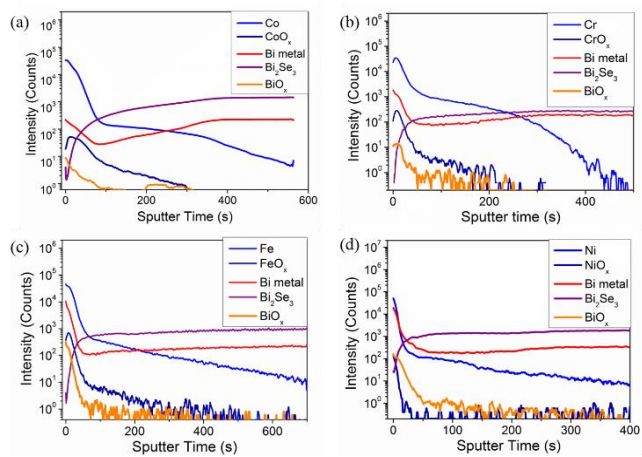

FIG. 3. SIMS results for samples coated with 1 nm of (a) Co, (b) Cr, (c) Fe and (d) Ni. (transition metal TM, TMOx, Bi metal and $Bi_2Se_3$ are reveled from measurements of $TM^+$, $TMO^+$, $Bi^+$ and $CsBi^+$ ions).

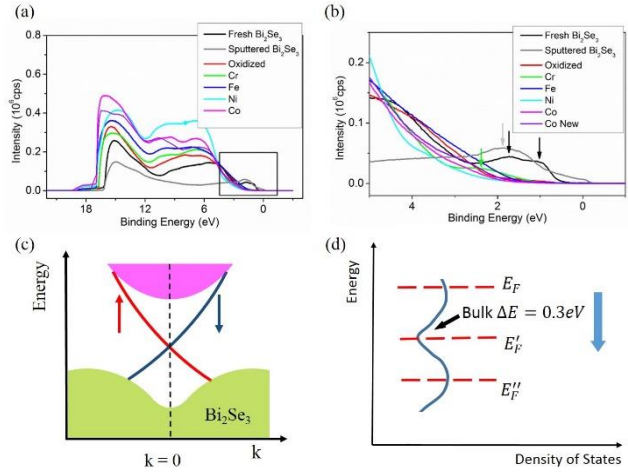

FIG. 4. (a) The UPS spectra; (b) the same zoomed in at small binding energies; (c) schematic of a band structure showing both the bulk and surface bands; (d) schematic density of states consistent with UPS results.

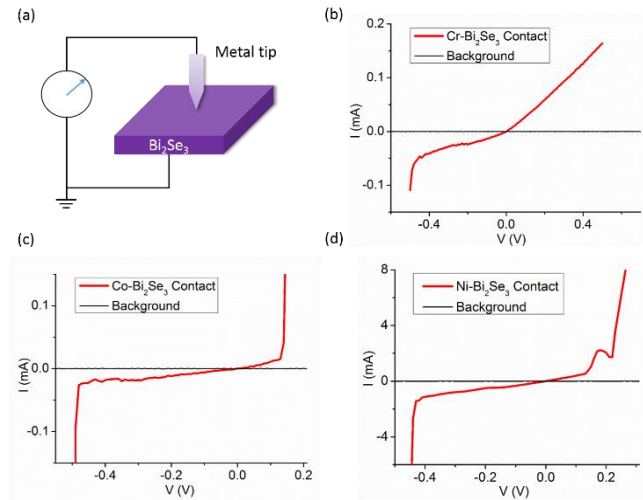

FIG. 5. (a) I-V measurement geometry; I-V curves for contacts with (b) Cr, (c) Co and (d) Ni;

[1] Joel E. Moore. *Nature* **464**, 194 (2010).

[2] M. Z. Hasan and C. L. Kane. *Rev. Mod. Phys.* **82**, 3045 (2010).

[3] Y. Ando. *J. Phys. Soc. Japan* **82**, 102001 (2013).

[4] Y.S. Hor, P. Roushan, H. Beidenkopf, J. Seo, D. Qu,2 J. G. Checkelsky, L. A. Wray, D. Hsieh, Y. Xia, S.-Y. Xu, D. Qian, M. Z. Hasan, N. P. Ong, A. Yazdani, and R. J. Cava, *Phys. Rev. B* **81**, 195203 (2010).

[5] Y. L. Chen, J.-H. Chu, J. G. Analytis, Z. K. Liu, K. Igarashi, H.-H. Kuo, X. L. Qi, S. K. Mo, R. G. Moore, D. H. Lu, M. Hashimoto, T. Sasagawa, S. C. Zhang, I. R. Fisher, Z. Hussain and Z. X. Shen. *Science* **329**, 659 (2010)

[6] J.G. Checkelsky, J. Ye, Y. Onose, Y. Iwasa and Y. Tokura. *Nat. Phys.* **8**, 729 (2012)

[7] A. R. Mellnik, J. S. Lee, A. Richardella, J.L.Grab, P. J. Mintun, M. H. Fischer, A.Vaezi, A.Manchon, E.-A.Kim, N. Samarth and D. C. Ralph. *Nature* **511**, 449–451 (24 July 2014).

[8] Y. Wang, P. Deorani, K. Banerjee, N. Koirala, M. Brahlek, S. Oh, and H. Yang. *Phys. Rev. Lett*. **114**, 257202 (2015).

[9] C. H. Li, O.M.J. van 't Erve, J.T. Robinson, Y. Liu, L. Li and B.T. Jonker, *Nature Nanotechnology* **9**, 218 (2014).

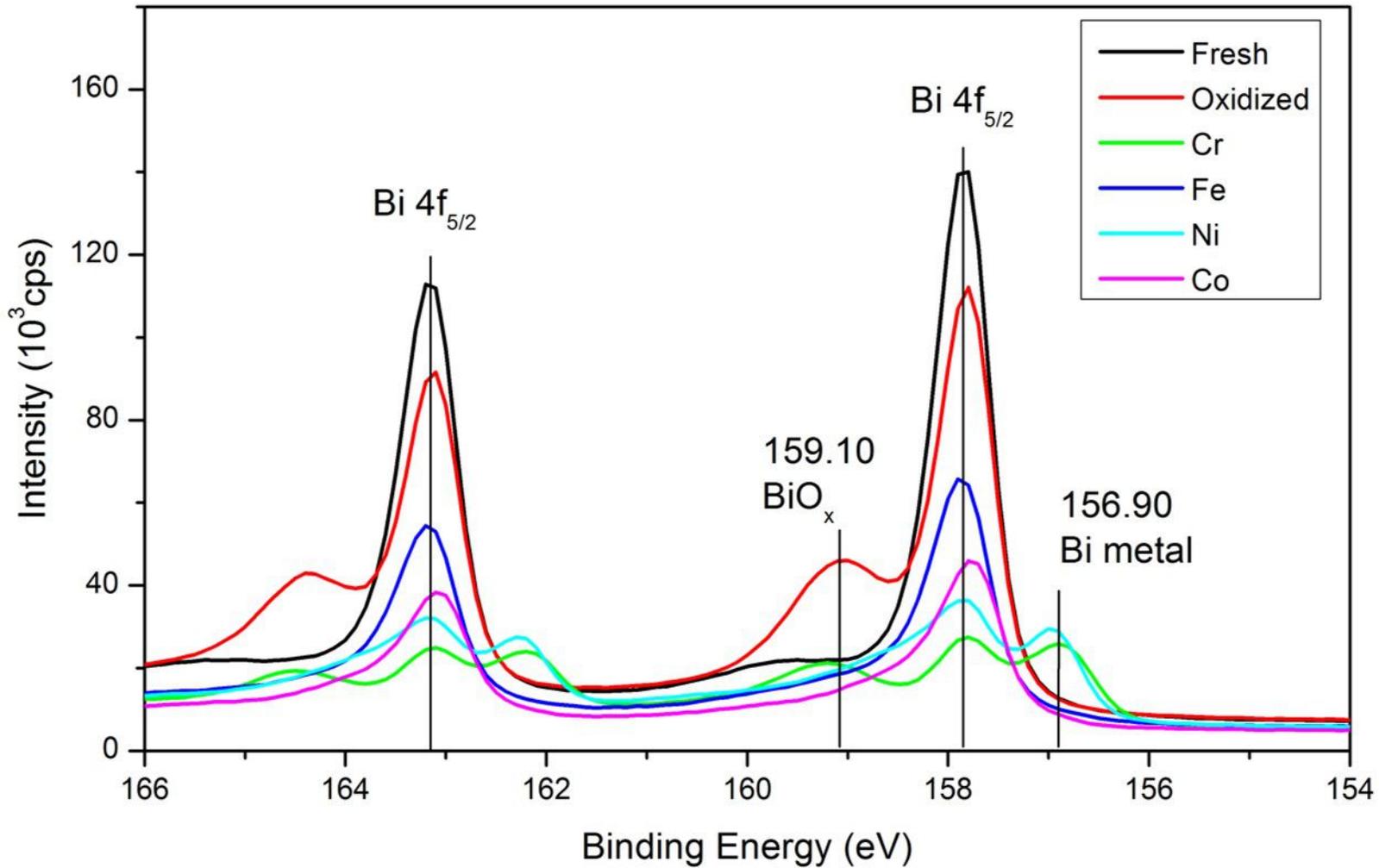

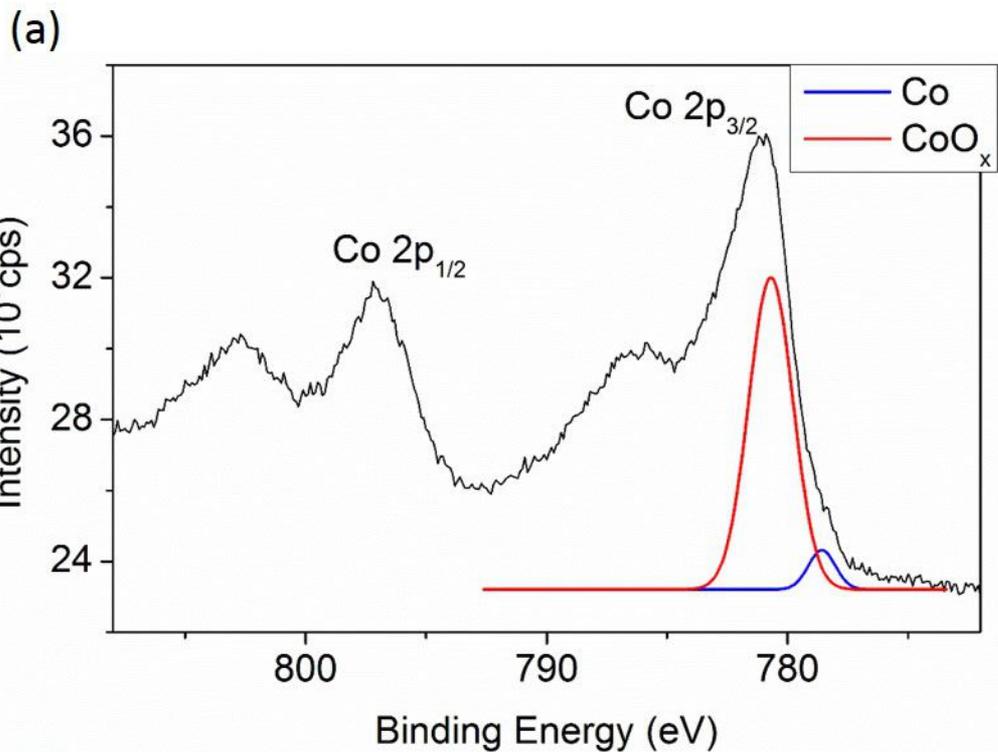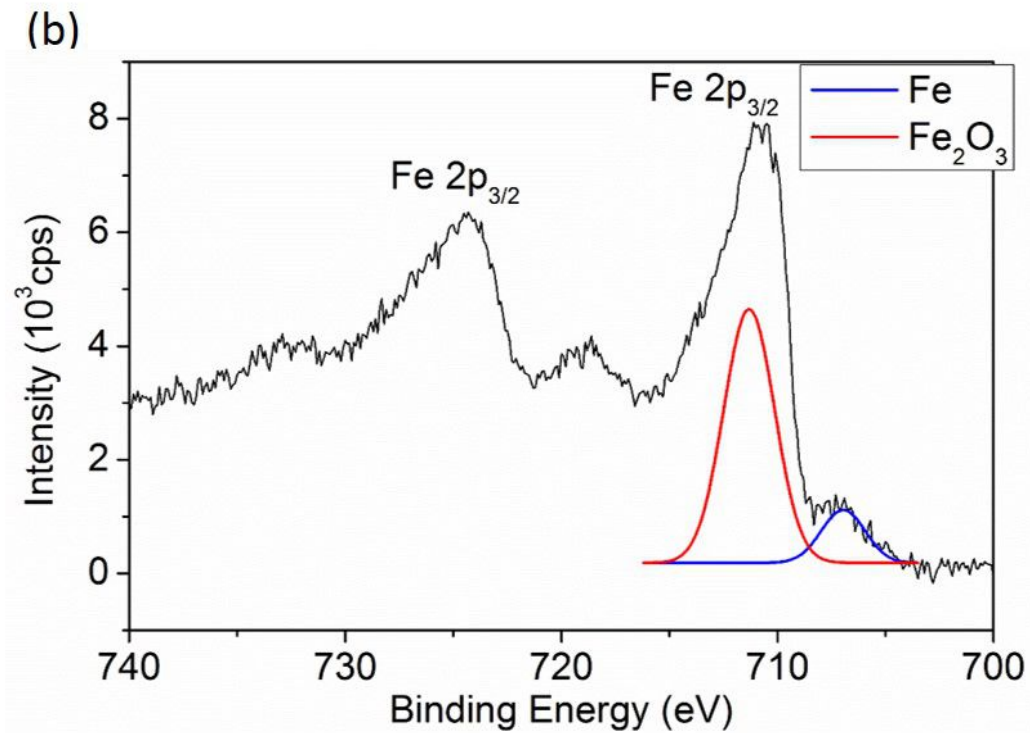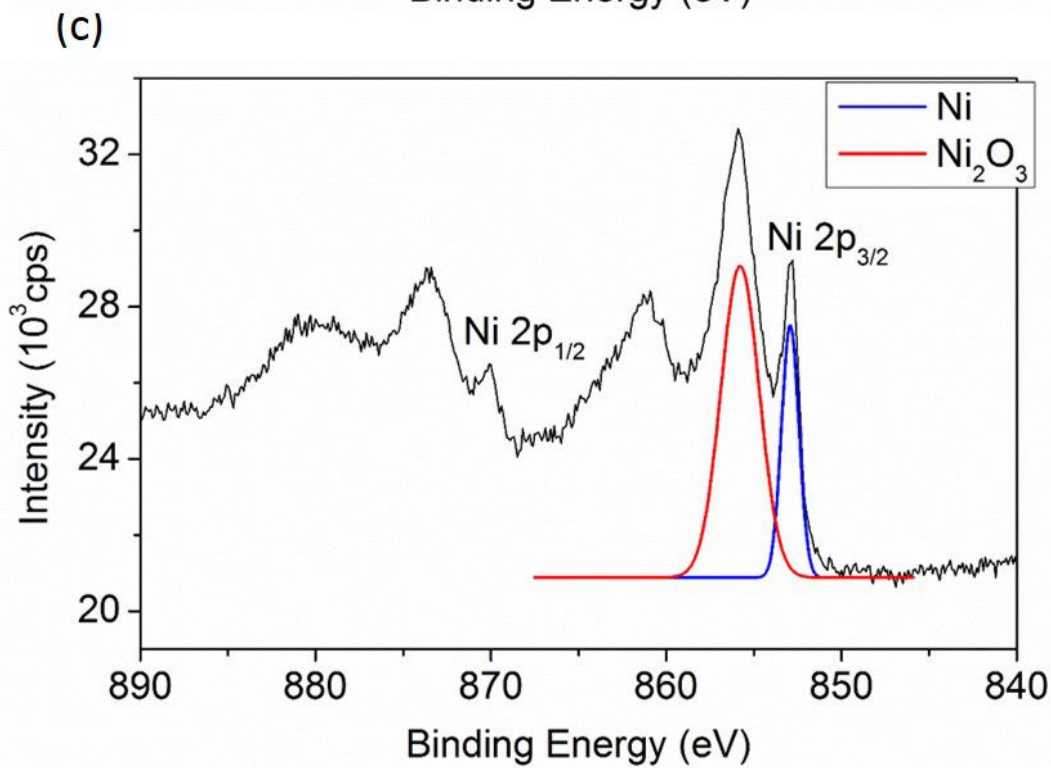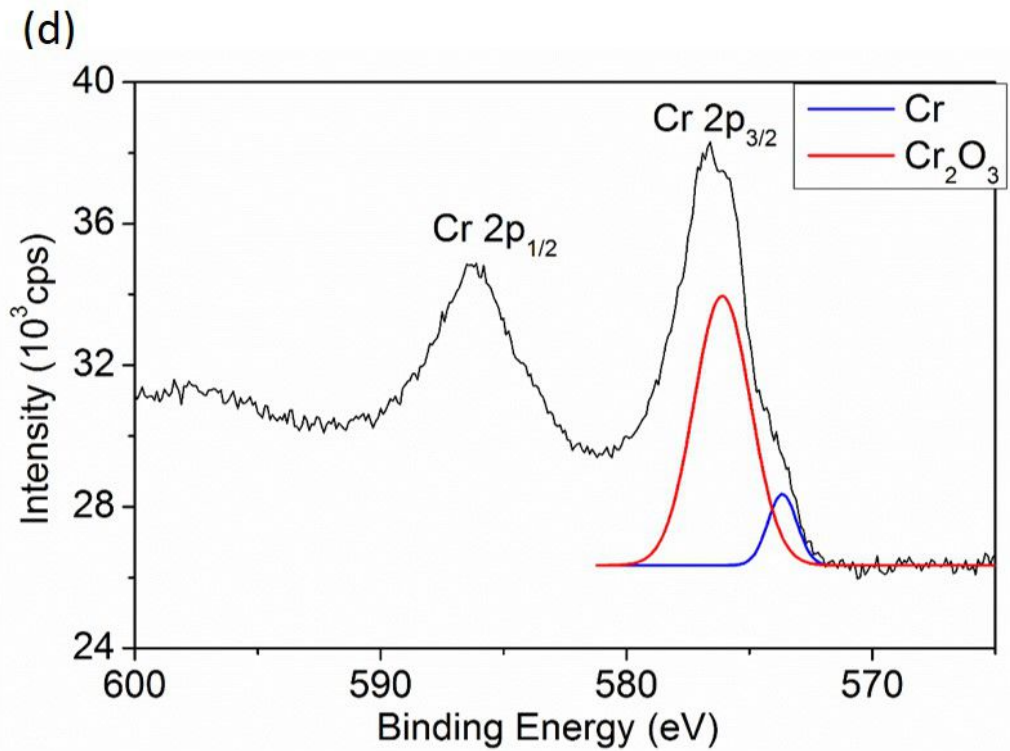

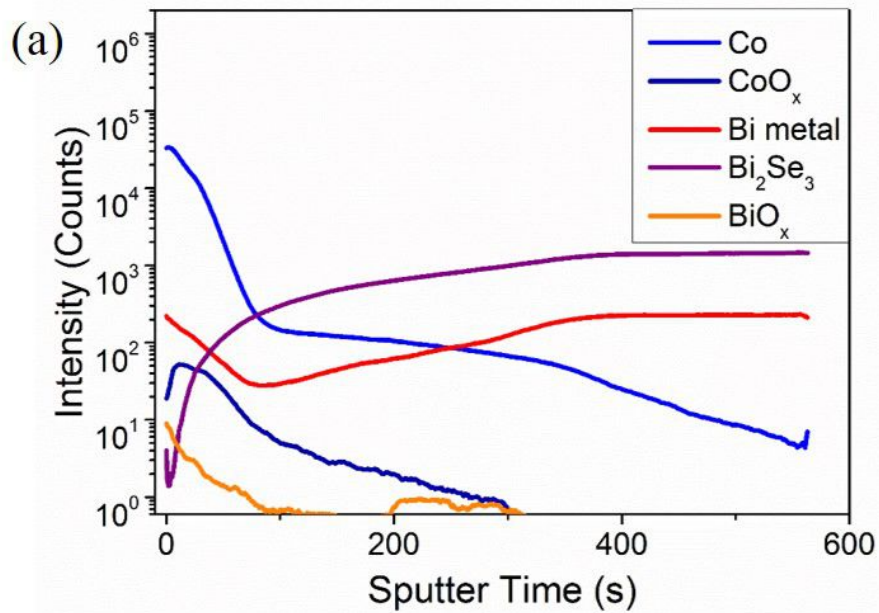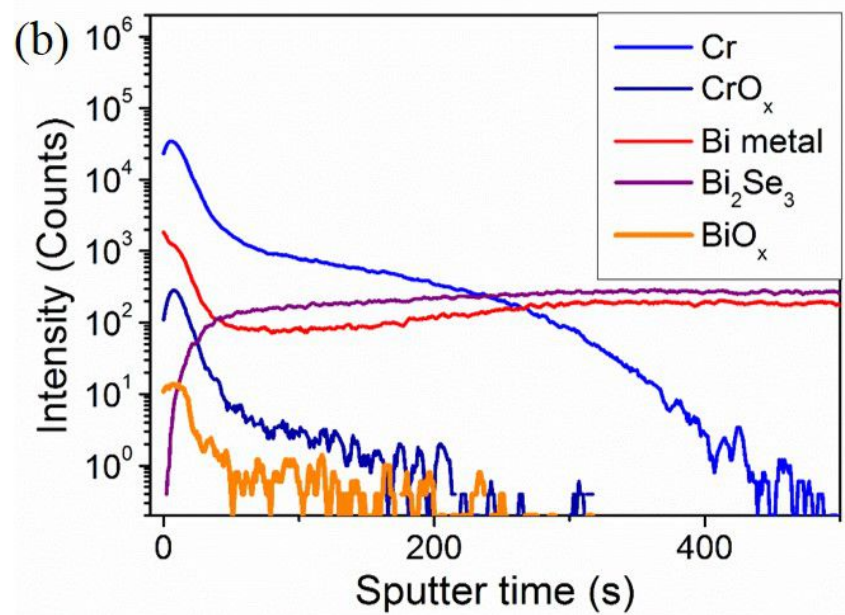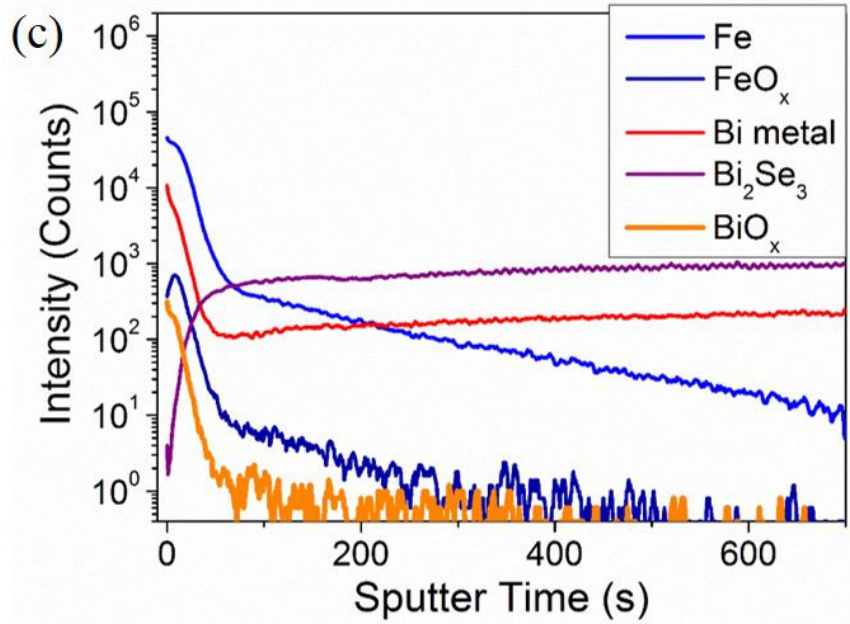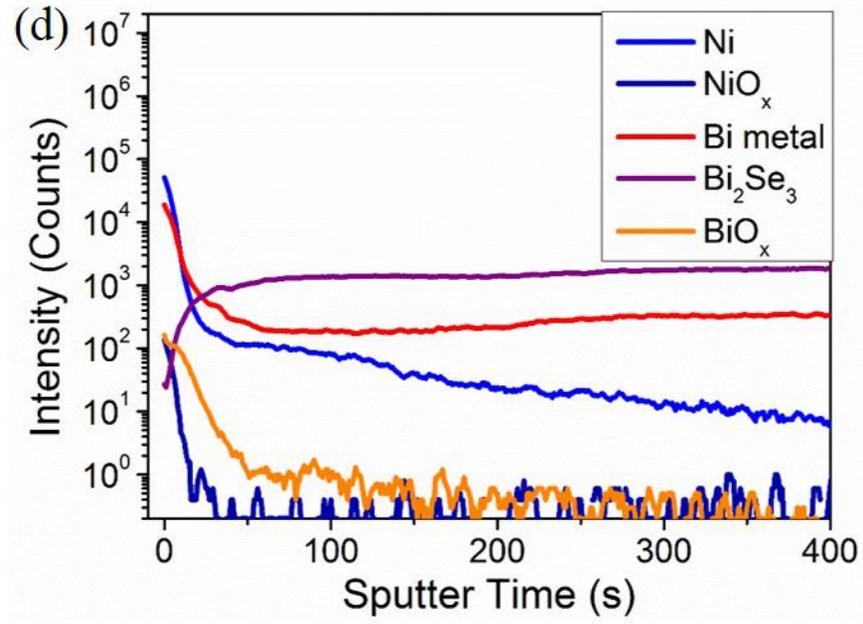

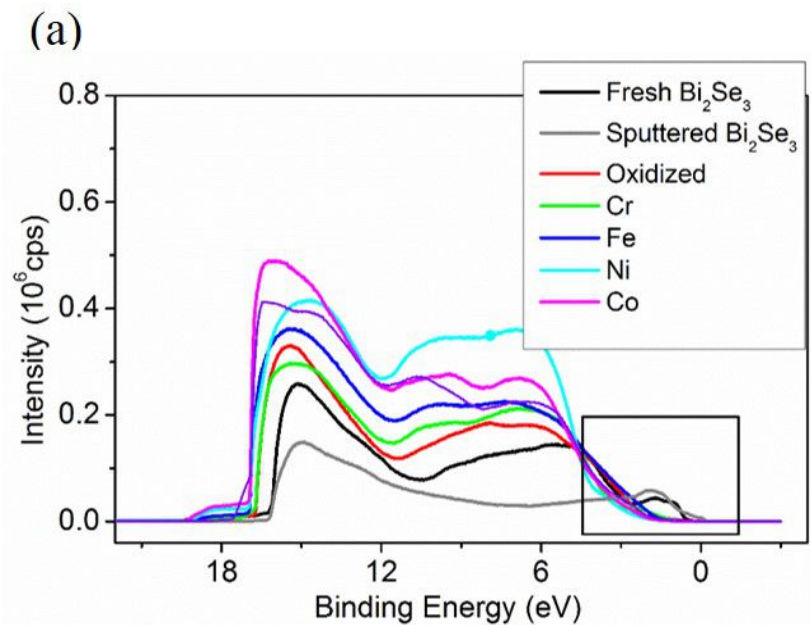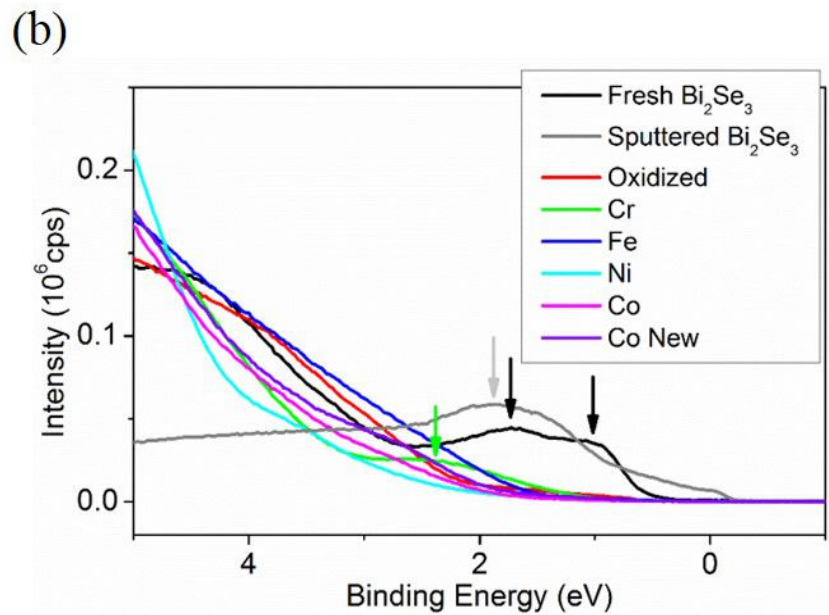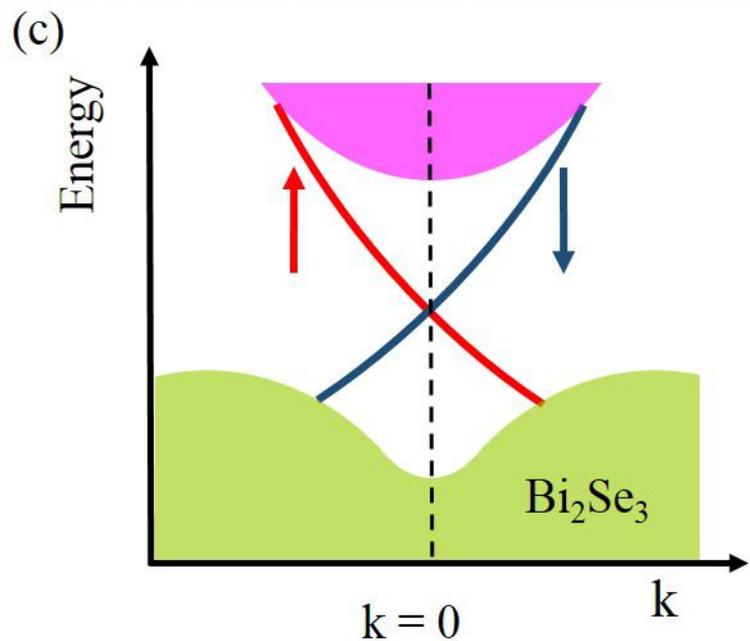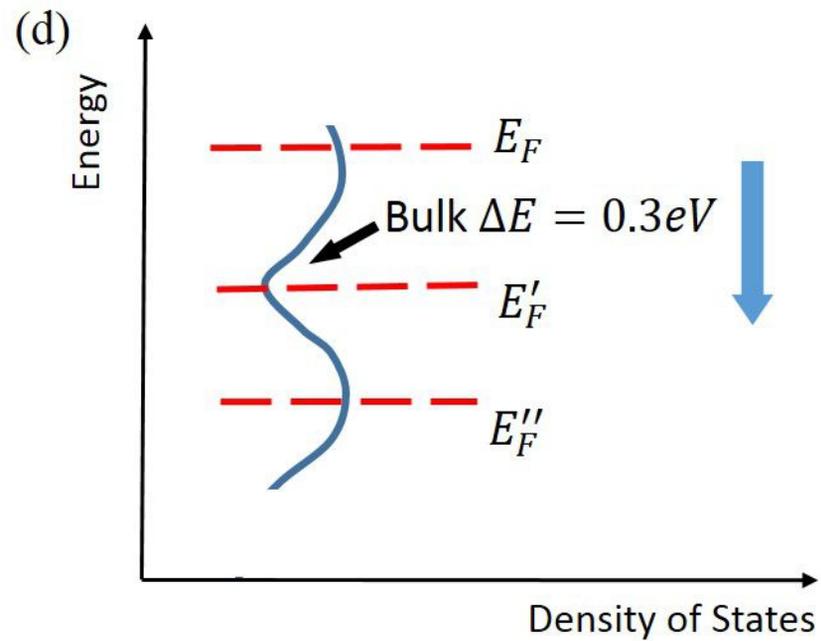

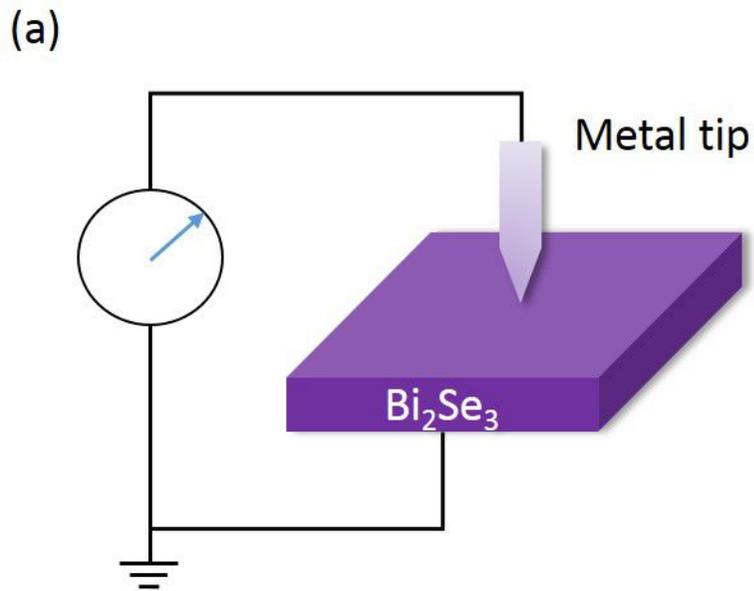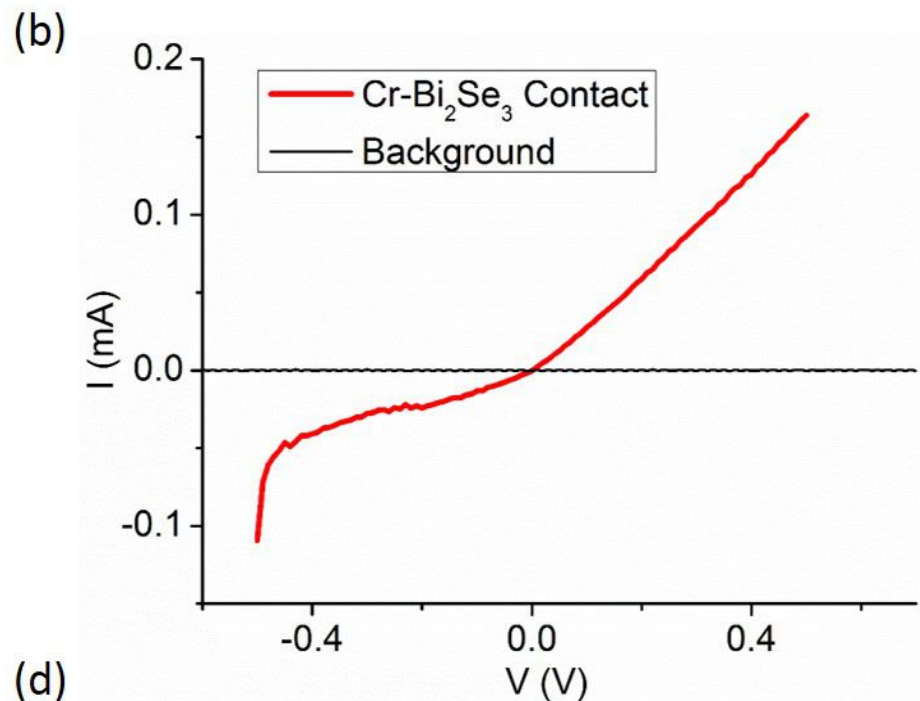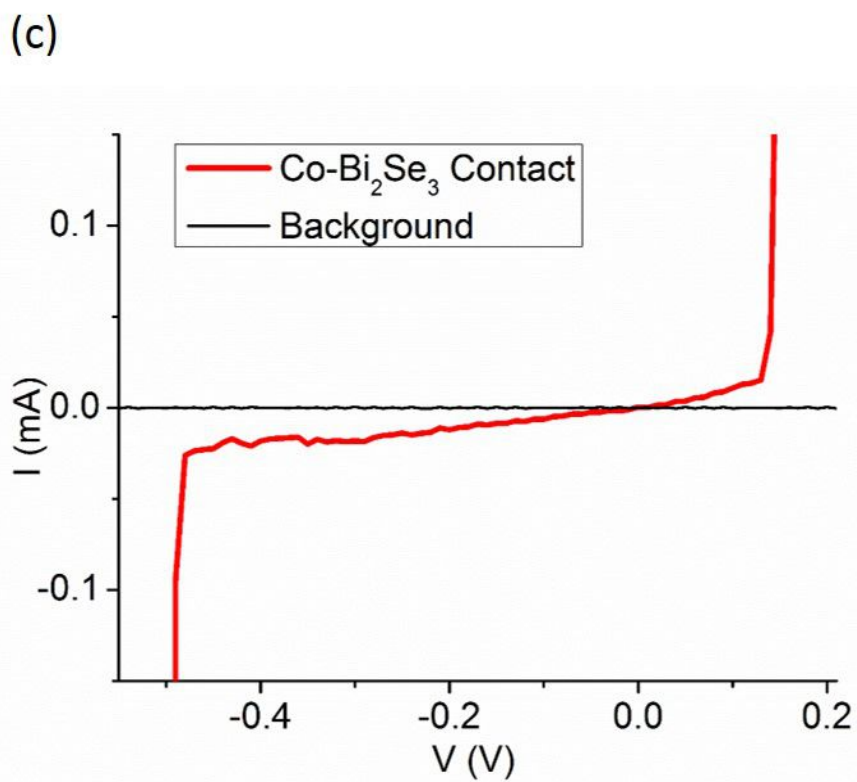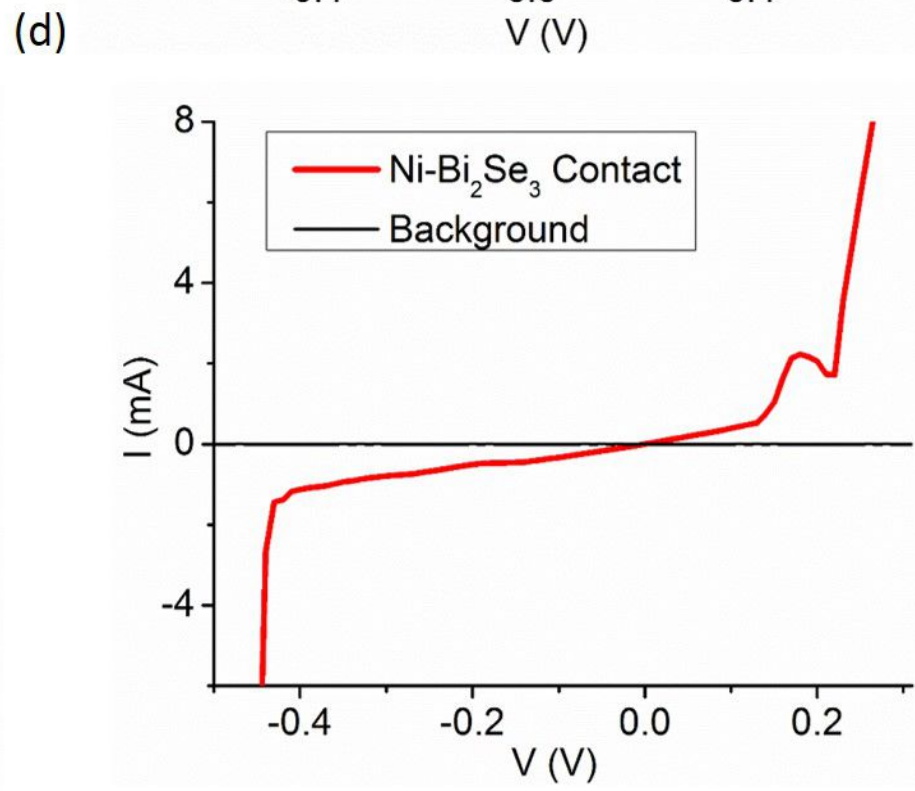